# Electrometer calibration with sub-part-per-million uncertainty

Hansjörg Scherer, Dietmar Drung, Christian Krause, Martin Götz, and Ulrich Becker

*Abstract*—We performed calibrations of four different commercial picoammeters using the Ultrastable Low-noise Current Amplifier (ULCA) as a calibrator current source operated in the range between 1 fA and 1 µA. The results allow the comprehensive characterization of the devices under test regarding noise, settling and burden voltage behavior as well as stability of the gain factor, and confirm the performance of the ULCA for use as small-current calibrator standard. Also, we present a further advanced setup for the calibration of transimpedance amplifiers. Accuracy limits for best electrometer calibrations in the current range between 1 fA and 1 µA and possible implications on corresponding calibration and measurement capabilities are discussed.

## I. Introduction

SMALL electrical currents and related instrument calibrations are relevant in various fields of research and application [1]. Lately, the Ultrastable Low-noise Current Amplifier (ULCA) was introduced by PTB as a powerful tabletop small-current instrument [2–4]. It enables the traceable measurement [5–7], but also the generation of small direct currents with unprecedented accuracy. Based on the novel transimpedance amplifier concept, different ULCA versions were developed: the 'standard' version with a total nominal transimpedance $A_{TR}$ = 1 GΩ and an input noise level of 2.4 fA/√Hz, described in [2–6], meanwhile is commercially available [8]; other ULCA variants with different features, tailored for special application purposes, are described in [7, 9–11]. Due to its performance regarding stability and versatility, the ULCA is highly suitable for the calibration of small-current amperemeters (picoammeters or electrometers) and sources, capable of outperforming other instruments and methods used for small-current calibrations in metrology [12, 13].

In this work, we applied PTB's calibrator ULCA, a prototype of the commercial 'standard' variant [8], as a current source standard for detailed investigations on four different commercial electrometer models. Besides the calibration of their gain factors, the ULCA-based calibration method enables the comprehensive characterization of relevant instrument parameters: noise and settling performance, gain factor stability as well as burden voltage behavior were investigated at highest accuracy levels. Characteristics and differences of the instruments are discussed.

Also, a further advanced setup for the calibration of electrometers of the kind "transimpedance amplifier" is described. We discuss the accuracy limits for calibrations in the current range between 1 fA and 1 µA as well as possible implications on corresponding calibration and measurement capabilities.

## II. Setup and Procedure

The setup used for the electrometer calibrations is shown in Fig. 1. A prototype of the 'standard' ULCA, operated in current source mode, was used as calibrator. The same instrument, shown in Fig. 2 of ref. [2], was also used for the measurements on single-electron pumps presented in [5, 6]. In the source mode (cf. Fig 3b in ref. [2]), a voltage $V_S$ is applied to the TEST input to generate a calibration current $I_S$ for the device under test (DUT) of up to ±5 nA. A servo-loop involving operational amplifier OA1 in Fig. 1 equalizes the internal ground potential to the burden voltage of the DUT, i.e., the voltage between the IN terminal and case. Therefore, the voltage at the ULCA output VOUT measured against internal ground with DVM1 is given by $V_{OUT} = -A_{TR} I_S$. The gain factor of the DUT is determined by the condition that the current measured by the DUT is identical to the calibration current $I_S$. Thus, although a possible burden voltage across the DUT affects $I_S$, it has no influence on the calibration result.

An AD/DA card was used as voltage source providing $V_S$ to bias the ULCA. As shown in Fig. 1, a low-pass filter (and, for low values of $I_S$, an additional resistive voltage divider) were used to suppress wide-band noise from the voltage source. The output voltage $V_{OUT}$ and the voltage $V_B$ were measured with two 8½ digit voltmeters, DVM1 and DVM2. An integration time of one power-line cycle was used, and an auto-zero process was manually triggered every two seconds.

The voltage $V_B$ measured between the GND terminal of the ULCA (being nominally at the same potential as the IN terminal due to the servo loop involving OA1) and the chassis ground is

This work was supported by the Joint Research Project 'e-SI-Amp' (15SIB08). It received funding from the European Metrology Programme for Innovation and Research (EMPIR), co-financed by the Participating States, and from the European Union's Horizon 2020 research and innovation programme.

H. Scherer (e-mail: hansjoerg.scherer@ptb.de), M. Götz, and U. Becker are with the Physikalisch-Technische Bundesanstalt (PTB), Bundesallee 100, 38116 Braunschweig, Germany. D. Drung and C. Krause are with the Physikalisch-Technische Bundesanstalt (PTB), Abbestraße 2-12, 10587 Berlin, Germany.

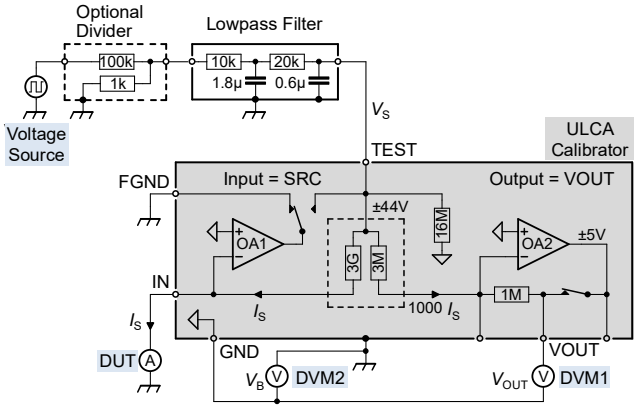

Fig. 1. Setup for the calibration of an electrometer (device under test, DUT) using an ULCA calibrator configured for current source mode ($I_S \leq 5$ nA).

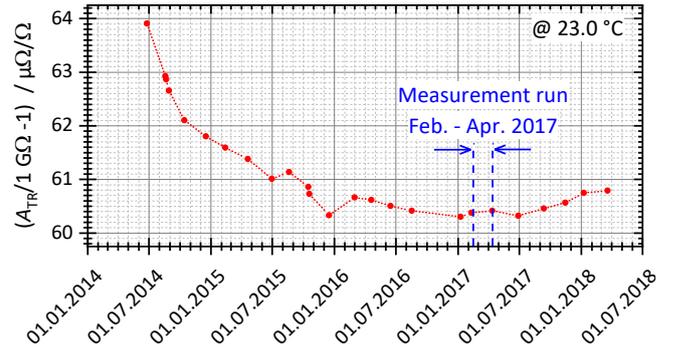

Fig. 2. Calibration results for the total transresistance $A_{TR}$ of the calibrator ULCA, all performed against a quantum Hall resistance standard and referred to a temperature of 23.0 °C. Error bars corresponding to the calibration uncertainty are smaller than the symbol size. The period of 7 weeks in which the measurements for section IV of this paper were performed is indicated.

the sum of the burden voltages of the calibrator source and the electrometer (DUT). Operation of the AD/DA card and digital data acquisition (DUT and DVM output signal reading) were controlled and synchronized by computer. The calibrator current $I_S$ applied to the DUT was reversed periodically every 500 s to suppress offset drifts. The connection between calibrator ULCA and DUT was established via a low-noise cable. The total source capacitance of < 200 pF including cable and ULCA does not cause any stability problems for the DUT instruments.

The calibrator ULCA was characterized by the typical 'standard' ULCA features: besides the low current noise of 2.4 fA/√Hz and the very low corner frequency of 1 mHz, the current settling is very fast: 3 s after polarity change, the deviation of $I_S$ from its final value is < 0.1 µA/A [2]. The temperature coefficient of $A_{TR}$ is -0.17 µΩ/Ω per kelvin. Before and after the measurements, $A_{TR}$ was calibrated traceable to a quantum Hall resistance (QHR) with a relative standard uncertainty of about 20 nΩ/Ω, using a cryogenic current comparator (CCC) according to the procedure described in [2] and [4]. Fig. 2 shows the calibration history of the calibrator ULCA monitored over a period of more than three years, with changes spanning over less than 4 µΩ/Ω. The relative change during the period of the measurement run for the results presented in section IV of this work (February to April 2017) was less than 0.1 µΩ/Ω.

All measurements were performed in a temperature stabilized laboratory, and their durations were chosen long enough to give significant results after data averaging. The polarity of the measurement currents ($I_S = \pm 5$ pA and $\pm 500$ pA) was reversed every 500 s to suppress offset drifts.

### III. ELECTROMETER DEVICES UNDER TEST

Regarding the four electrometers investigated in this paper, two instrument categories can be distinguished: digital picoammeters, and analog transimpedance amplifiers with analog voltage output. For objective comparability, all instruments in this study were used as delivered by the manufacturer, and with original auxiliary equipment.

#### A. Digital Picoammeters

Two of the commercial electrometers investigated were digital instruments[1]: DUT1 was a 'Sub-Femtoamp Remote SourceMeter' (Keithley Model 6430), DUT2 was a 'Femto/Picoammeter and Electrometer/High Resistance Meter' (Keysight Model B2985A).

Both DUTs were set up in current metering mode using the lowest input current ranges for measurements at different current levels: for the measurements at $I_S = \pm 5$ pA this was the 10 pA (20 pA) range for DUT1 (DUT2), while for the measurements at $I_S = \pm 500$ pA the range chosen was 1 nA (2 nA). Internal filters of the instruments were deactivated during the measurements to provide comparable results. DUT signals were directly read out via their GPIB computer interface.

#### B. Transimpedance Amplifiers

Two commercially available analog transimpedance amplifiers[1] were investigated: DUT3 was 'Variable Gain Sub-Femtoampere Amplifier' (Femto Model DDPCA-300), and DUT4 was the commercialized 'standard' version of the ULCA (Magnicon Model ULCA-1 [8]) with a fixed total transimpedance of 1 GΩ.

The variable gain of DUT3 was set to 1 TΩ (10 GΩ) transimpedance for the measurements at ±5 pA (±500 pA). To deactivate internal low-pass filters, the bandwidth was set to "full" (rise time "fast"). Also, the original power supply was used with DUT3. DUT4 was configured for current measurement mode as described in Fig. 3a of ref. [2]. Output voltages of both instruments were digitized with an additional 8½ digit DVM not shown in Fig. 1.

---

[1] The identification of specific commercial instruments does not imply endorsement by PTB nor does it imply that the instruments identified are the best available for a particular purpose.

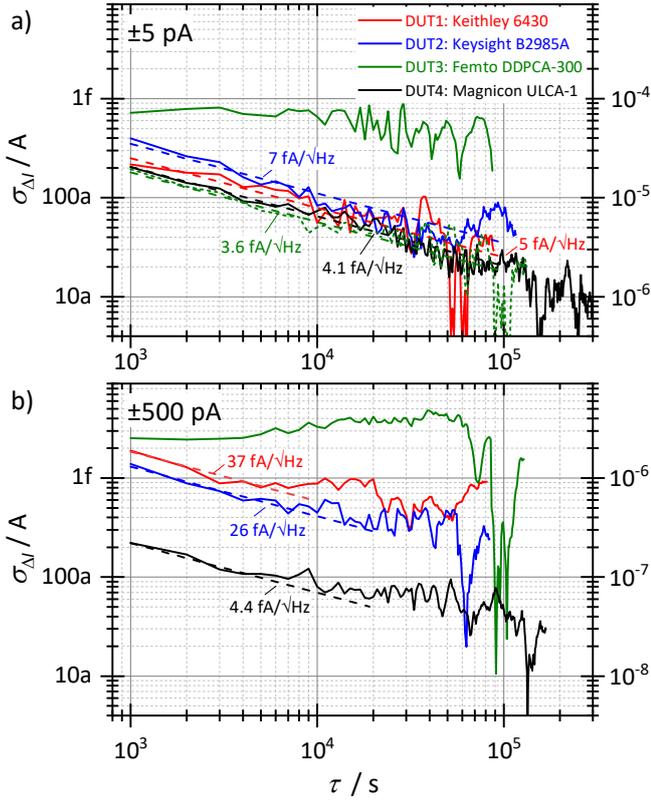

Fig. 3. Allan deviation plots of the DUT error current $\Delta I$ for $I_S = \pm 5$ pA and $\pm 500$ pA. Current values as input data for the analysis were computed from the differences of DUT signal data averaged over the negative and positive half-cycles. The first 50 s after each current reversal were rejected to suppress transients. White noise levels, indicated by the dotted fit lines, were calculated using Eq. 2 in ref. [2], considering an effective integration time $\tau_e = 0.9 \cdot \tau$. The axes on the right side are scaled in relative units, normalized to $I_{pp} = 2|I_S| = 10$ pA and 1 nA, respectively. Panel a) shows an additional measurement on DUT3 performed at $\pm 50$ fA (green dotted curve), which corresponds to a noise level of 3.6 fA/√Hz. In this case, the right scale is to be multiplied by a factor of 100 to account for the reduced current level.

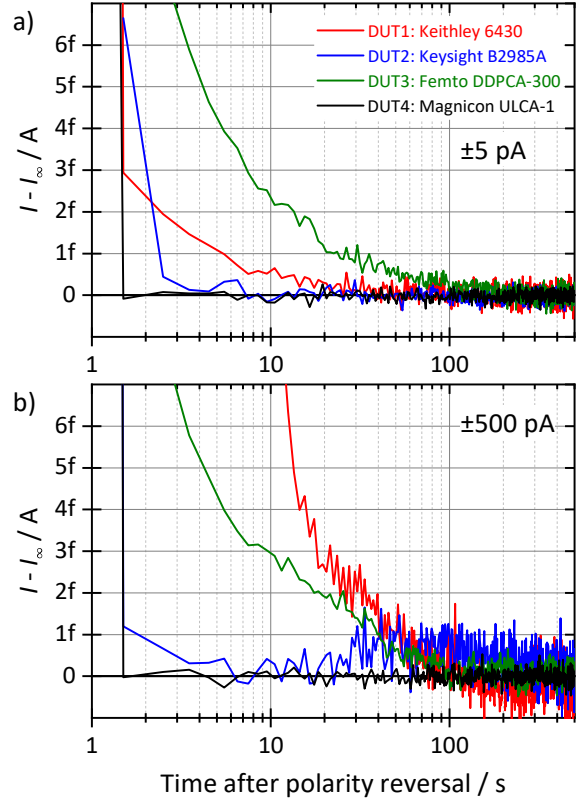

Fig. 4. Settling behavior of the DUTs. Plotted are the DUT output signals versus time after polarity reversal for $I_S = \pm 5$ pA and $\pm 500$ pA. $I_\infty$ is the final current signal asymptotically approached. Each curve is the result from averaged data measured over periods of about two to four days. Responses from both switching directions were considered: data from positive current steps were directly used, whereas those for negative steps were inverted before averaging.

## IV. RESULTS

This section presents the measurement results, analyzed to compare the four electrometer instruments regarding their current noise, settling and burden voltage behavior.

### A. Current Fluctuations

Fig. 3 shows Allan deviation plots of the DUT error current $\Delta I$ (i.e., the difference between the currents displayed by the DUT and the calibrator ULCA) for integration times longer than $\tau = 1000$ s. Generally, at sufficiently low current levels, the Allan deviation $\sigma_{\Delta I}$ is dominated by amplifier noise. In this regime, it decreases with the square root of the sampling time $\tau$ due to periodic current reversal [4]. For larger currents, low-frequency fluctuations in the transresistance gain increase the Allan deviation. This effect is only weak for DUT4 due to its high gain stability. Given the known current noise level of 2.4 fA/√Hz of DUT4, a total effective current noise of 3.8 fA/√Hz is estimated for the calibrator setup including the DVM contribution, in reasonable agreement with the measurement results. In case of DUT3, the Allan deviation for both current levels ±5 pA and ±500 pA is rather high and roughly constant over the time scale of Fig. 3. This is presumably caused by the thick-film resistors used in the feedback of the input amplifier. For testing this assumption, we reduced the current level from ±5 pA to ±50 fA and found a noise level of 3.6 fA/√Hz including the noise from the ULCA and both DVMs (green dotted curve in Fig. 3a), presumably caused 1/f amplifier noise from DUT3. According to the manufacturer specifications for DUT3, noise levels of 0.2 fA/√Hz at 0.4 Hz (1.3 fA/√Hz at 1 Hz) for the 1 TΩ (10 GΩ) gain setting are expected. These noise levels were verified by test measurements with open input (zero input current, not shown in Fig. 3).

### B. Settling

Fig. 4 shows the settling behavior of the DUTs for $I_S = \pm 5$ pA and $\pm 500$ pA. DUT1 and DUT3 settle relatively slow, and the performance strongly depends on the current range settings. According to the manufacturer specification for DUT2, this electrometer type was designed for fast settling times, which is reflected in the results shown in Fig. 4. In comparison, DUT4 shows the fastest settling at both current levels. As reported in

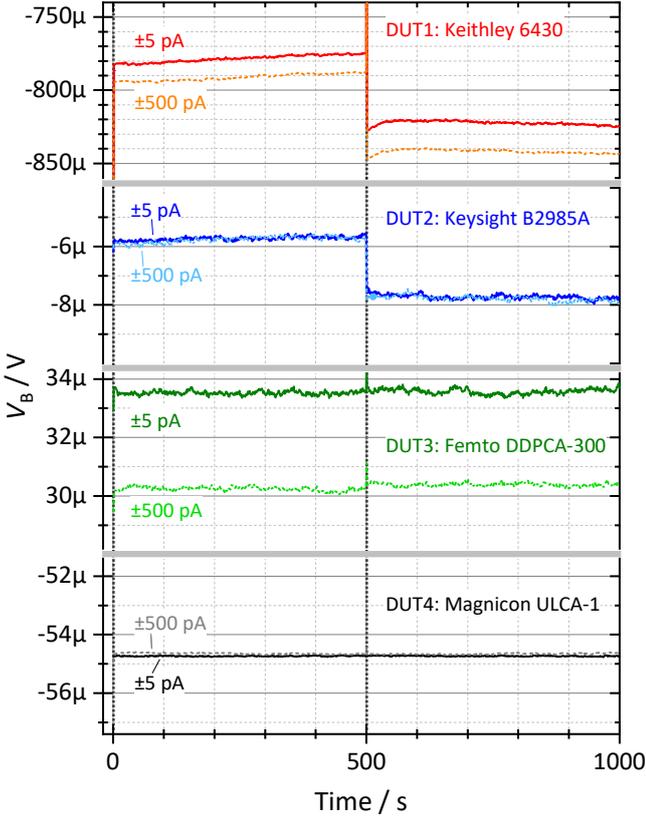

Fig. 5. Measurements of $V_B$ versus time for $I_S = \pm 5$ pA and $\pm 500$ pA. Current polarity reversal happened at $t = 0$ s and 500 s. Note that the range of the y-axis for DUT1 is 20 times the range for the other DUTs. Each curve is the result from averaged data measured over a period of about two to four days.

[2], this instrument typically approaches its final value to within 0.1 µA/A only 3 s after polarity change of $I_S$.

*C. Burden Voltage*

Fig. 5 shows the results of $V_B$ measurements for $I_S = \pm 5$ pA and $\pm 500$ pA. As explained in section II, the voltage $V_B$ measured in the setup according to Fig. 1 is the sum of the burden voltages of the ULCA calibrator and the electrometer under test. We first discuss the levels of the burden voltages measured. DUT1 shows comparably large burden voltages of about 800 µV. This is, however, in agreement with the manufacturer specifications, claiming an upper burden voltage limit of 1 mV for this instrument. Lower burden voltages in the µV range were observed on DUT2. This agrees with the manufacturer specifications, claiming maximum input burden voltages of 20 µV for both current ranges used. Slightly higher burden voltages of about 30 µV were measured on DUT3. For DUT4, burden voltages of about -55 µV were observed.

More important than the absolute level of burden voltage is its change upon current reversal, also shown in Fig. 5: DUT1 shows a relatively large change of $V_B$ of about 50 µV together with peak-like overshoot. Also, pronounced transients over periods longer than 100 s were detected. This behavior can be explained by limited open-loop gain of the amplifier stage. For DUT2 a smaller, step-like change of the order of 2 µV was observed. For DUT3, the change of $V_B$ after current reversal is well below 1 µV. For DUT4, no burden voltage changes were detectable on the sub-microvolt scale. This is due to the high open-loop gain of well above $10^9$, which again is due to the specially designed operational amplifiers in the ULCA [2–4].

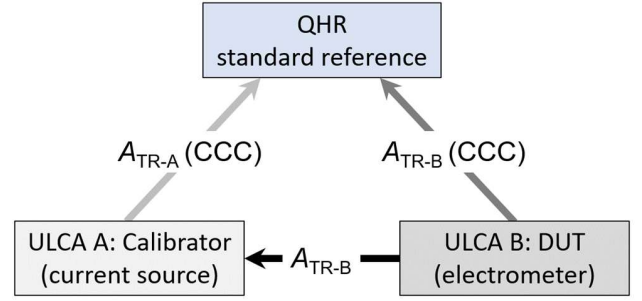

Fig. 6. Triangle comparison involving two ULCA instruments A and B, both calibrated against QHR with highest accuracy and directly compared using the calibration setup shown in Fig. 1.

Finally, for a comparison of the different electrometers with respect to their practical applicability, their temperature coefficients also need to be considered: for the two digital picoammeters DUT1 and DUT2 investigated, temperature coefficients of about 100 µA/A per kelvin are specified by the manufacturers. For the transimpedance amplifier with variable gain, DUT3, the specification is about 300 µA/A per kelvin. Such values are typical for amplifier stages involving thick-film feedback resistors. Electrometers of type ULCA-1 (DUT4), however, typically show temperature coefficients for $A_{TR}$ of less than 1 µΩ/Ω per kelvin [2]. This low value is enabled by the specially designed thin-film resistor network in the current amplification stage, and the metal-foil resistors in the current-to-voltage converter stage.

V. UNCERTAINTY LIMITS

Criteria for the determination of calibration and measurement capabilities (CMCs) are given in Ref. [14]. Here, on page 2 under N5 the following rule is formulated: "*Contributions to the uncertainty stated on the calibration certificate include the measured performance of the device under test during its calibration at the NMI or accredited laboratory. CMC uncertainty statements anticipate this situation by incorporating agreed-upon values for the best existing devices.*" Regarding the results presented in section IV and considering the unparalleled long-term stability of its transresistance, the ULCA-1 [8] can be considered the "best existing device" in the picoampere range, and is expected to have implications on future determinations of CMCs in the field of small currents.

To evaluate the uncertainty limits for calibrations with the setup shown in Fig. 1 (i.e. in the configuration "ULCA calibrates ULCA"), we performed a triangle comparison as sketched in Fig. 6. First, both ULCAs were calibrated traceable to QHR using PTB's 14-bit CCC, yielding the transresistance values $A_{TR-A}$ and $A_{TR-B}$ with relative standard uncertainties < 0.02 µΩ/Ω. Next, the transresistance values of both ULCAs were directly compared at different currents between 1 fA and

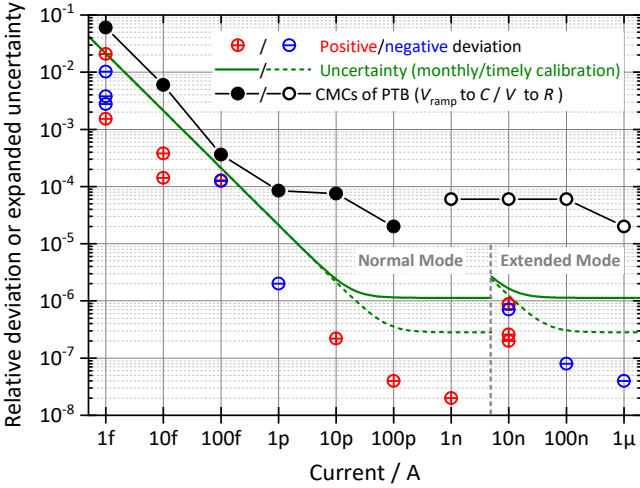

Fig. 7. Red and blue circled symbols show the measurement results from the triangle comparison as sketched in Fig. 6, plotted as relative deviations $\Delta A_{TR}/1\,G\Omega$ (+/- symbols indicate positive/negative values). Green lines show calculated expected upper limits for the measurement results, based on uncertainty estimates. Black circle symbols represent the present CMC values of PTB for two different, not ULCA-based calibration methods. "Normal" ("extended") mode refers to the current $I_S < (>)$ 5 nA and corresponding configurations of the ULCA instruments. All values correspond to expanded uncertainties (coverage factor $k = 2$, or 95% confidence interval).

1 µA using the calibration setup shown in Fig. 1.

Results from this experiment, performed with the ULCA calibrator source introduced in section II and a commercial 'standard' ULCA (model ULCA-1 from Magnicon [8]), are shown in Fig. 7. The red and blue circled symbols show measurement results for deviations of $A_{TR-B}$ determined by "ULCA-ULCA" comparison from the expectation value of $A_{TR-B}$ determined by CCC calibration: $\Delta A_{TR} = A_{TR-B} - A_{TR-B}(CCC)$. Positive (negative) results for $\Delta A_{TR}$ are indicated by "+" ("-") symbols. For $I_S < (>)$ 5 nA, the ULCA instruments were configured for "normal" ("extended") mode operation according to [9]. The DVM instruments reading $V_{OUT}$ on ULCA A and B were set to 100 mV range except for 1 nA and 1 µA, where the 1 V range was used. For $I_S \leq 100$ pA, measurement durations per point were about 20 h, while for larger currents each measurement took 67 min.

Green lines in Fig. 7 represent uncertainty estimates derived from calculations based on uncertainty figures of the setup components according to the following equation for the total standard uncertainty $u_{comb}$ ($k$=1):

$$u_{comb} = \sqrt{2\left(\frac{2S_I}{I_{pp}^2 t_{eff}} + u_{syst}^2\right)} \quad (1)$$

Here, the first term in the sum under the square root considers the statistical uncertainty contribution according to Eq. (2) in Ref. [2]. For "normal" ("extended") operation mode, $\sqrt{S_I}$ = 2.7 fA/√Hz (360 fA/√Hz) effective current noise was considered per ULCA including DVM, and the division by the by the peak-to-peak current value $I_{pp}$ accounts for the uncertainty given in relative units. Measurement times as stated above with 90% of data usage are assumed. The systematic contribution $u_{syst}$ in Eq. (1) is mainly caused by uncertainties in the calibration of the DVMs and the ULCAs. Two cases are considered in Fig. 7: monthly and timely calibration, the latter implying that the measurement is taken within one day before or after calibration. For the gain factor of each voltmeter and for timely calibration, a relative contribution of 0.08 µV/V was attributed, based on an estimate following typical results from calibrations with a Josephson voltage standard [15]. The drift in the gain factor 0.6 µV/V per month, as estimated from Ref. [15], was considered by an uncertainty contribution of 0.35 µV/V with uniform distribution. For the transresistance of each ULCA, a relative uncertainty of 0.04 µΩ/Ω [9] for timely calibration and a drift of 0.2 µΩ/Ω per month [18] were assumed. All other uncertainty contributions are much smaller and are included in $u_{syst}$ = 0.1 µA/A used for timely calibration or $u_{syst}$ = 0.4 µA/A used for monthly calibration.

The factor of 2 under the square root symbol in Eq. (1) accounts for the uncertainty contributions from two ULCA and two DVM instruments.

The green lines in Fig. 7 show that for monthly calibration, statistical contributions (noise) dominate the uncertainties for currents below about 10 pA, while for higher currents the uncertainty limit is dominated by systematic contributions from the DVM calibrations. As expected, the uncertainty estimations corresponding to the green lines in Fig. 7 represent upper limits for the measurement results (red and blue symbols).

Black circle symbols represent the present CMC values of PTB [16] for two different, not ULCA-based calibration methods: the "voltage-ramp-to-capacitor" method for currents < 100 pA [17], and the "voltage-to-shunt resistor" method for higher currents.

Altogether, these results verify the ULCA as small-current source calibrator and show that the calibration method based on the setup shown in Fig. 1 sets new benchmarks for the accuracy limits for electrometer calibrations in the current range between 1 pA and 1 µA. Potential impact on corresponding calibration and measurement capabilities can be derived from a comparison with PTB's present CMCs [16]. For currents above a few picoamperes, for instance, improvements of more than a factor of 10 in uncertainty seem realistic.

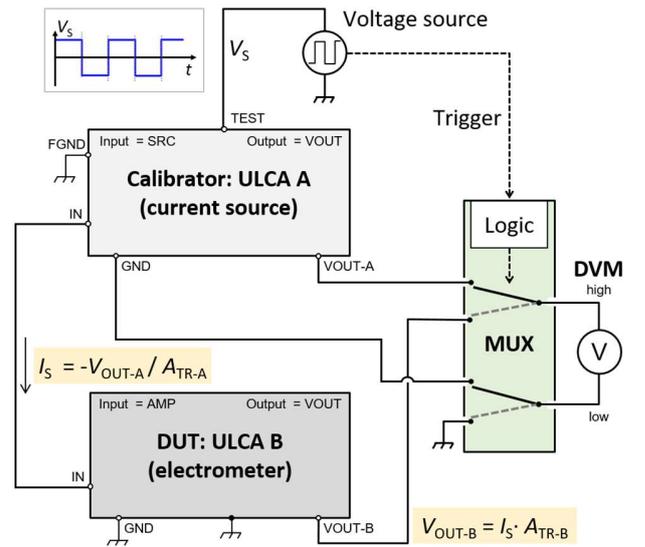

Fig. 8. Advanced setup for the calibration of electrometers with analog voltage output (here: ULCA B) with ULCA A used as calibrator current source, involving a multiplexer circuit (MUX) and a single DVM to measure the outputs of both instruments alternatingly. The voltage source biasing the calibrator ULCA with $V_S$ also gives the trigger signal controlling the MUX switching events.

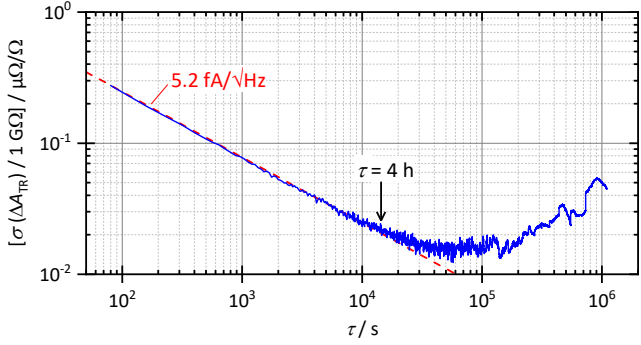

Fig. 9. Allan deviation plot of the ULCA-ULCA calibration measurements. The red dotted line is a fit corresponding to white noise of 5.2 fA/√Hz, calculated using Eq. 2 in ref. [2] with an effective integration time $\tau_e = \tau/4$. $\tau_e$ is reduced with respect to $\tau$ by a factor of 2 due to the sequential measurement mode (using the MUX), and by another factor of 2 due to the disregarded data after current reversal (suppressing transients).

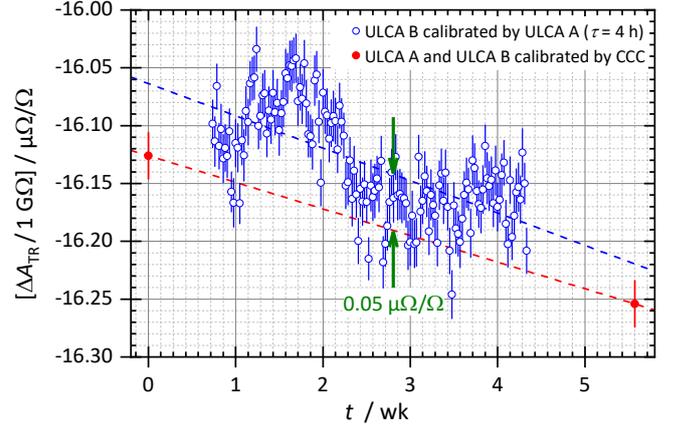

Fig. 10. Blue data points (open symbols) show calibration results using the setup shown in Fig. 8 with $I_S = \pm 3$ nA and data averaged for $\tau = 4$ h. Red points (full symbols) show the results from calibrations of $A_{TR-A}$ and $A_{TR-B}$ performed against QHR by using a cryogenic current comparator. Dotted lines represent linear fits to both data sets.

## VI. ADVANCED CALIBRATION SETUP

As explained in sections II and III, the setup shown in Fig. 1 needs two DVMs for gain calibrations of electrometers with analog voltage output: the first (DVM1 in Fig. 1) is needed to measure the output voltage of the ULCA calibrator, further called $V_{OUT-A}$ according to the denomination introduced in Fig. 6. The calibrator current $I_S = -V_{OUT-A}/A_{TR-A}$ is sourced to the electrometer with transresistance (gain) $A_{TR-B}$, which performs current-to-voltage conversion resulting in an output voltage $V_{OUT-B} = I_S \cdot A_{TR-B}$ to be measured with a second DVM (not shown in Fig. 1). Given the known calibrator transresistance $A_{TR-A}$, the gain factor $A_{TR-B}$ to be determined is derived from $A_{TR-B} = -A_{TR-A} \cdot V_{OUT-B}/V_{OUT-A}$. Correspondingly, the gain uncertainties of both DVMs take influence on the calibration result.

A further advanced setup avoiding this drawback is shown in Fig. 8. Here, a single DVM is used for measuring the output voltages of the calibrator source and of the electrometer under calibration by using a multiplexer unit (MUX). The MUX interchanges the DVM between the calibrator and the electrometer periodically and synchronized with the current polarity reversals. Since the calibration result is calculated from the ratio of $V_{OUT-B}$ and $V_{OUT-A}$, uncertainties of the DVM gain and linearity are suppressed. Note, however, that this sequential measurement mode reduces the effective integration time by a factor of 2 ($\tau_e = \tau/2$).

Generally, this setup can be used for the calibration of transimpedance amplifiers with transimpedance levels close to the calibrator instrument (i.e. 1 GΩ for the for the 'standard' ULCA). To assume the best possible configuration, but without loss of generality, here an ULCA was chosen as electrometer to be calibrated.

Figs. 9 and 10 shows results from such "ULCA-ULCA" calibration measurements performed over a period of about four weeks, with $I_S = \pm 3$ nA being reversed every 10 s and the first 5 s after each reversal being disregarded to reject transients. The DVM was switched by the multiplexer between ULCA A and ULCA B synchronously with the current reversals. For both plots, the relative difference $\Delta A_{TR} = A_{TR-A} - A_{TR-B}$ normalized to 1 GΩ was evaluated.

In Fig. 9, the Allan deviation of $\Delta A_{TR}/1$ GΩ is plotted. The total effective white noise level of 5.2 fA/√Hz in this measurement is higher than the combined noise level of 3.4 fA/√Hz stemming from the two ULCAs. The excess noise is presumably caused by the DVM and the voltage source. It is noted that the effective noise level of this setup is comparable to the state-of-the-art CCC-based calibration setup presented in [4]. For integration times $\tau > 4$ h, transition from the white noise to the flicker noise regime becomes visible.

In Fig. 10, averaged data (blue points) from 4 h long measurement intervals are plotted versus time. In agreement with Fig. 9, the time series shows typical flicker noise characteristics. Within one day, fluctuations of $\Delta A_{TR}/1$ GΩ up to about 0.05 μΩ/Ω were observed. Fig. 10 also shows the results from the calibration measurements performed with the CCC traceable to QHR before and after the "ULCA-ULCA" measurements (red points). Dotted lines represent linear fits to both data sets. In comparison, they show an agreement of better than 0.1 μΩ/Ω and indicate a change of $\Delta A_{TR}/1$ GΩ of about -0.025 μΩ/Ω per week.

Altogether, these results document that the advanced setup shown in Fig. 8 enables "ULCA-ULCA" calibrations with an accuracy of about 0.04 μΩ/Ω in 1 h of measurement time. This comes close to the calibration uncertainty level reachable with PTB's CCC setup [4]. The advanced setup is suitable for laboratories that are equipped with several ULCA instruments but lacking a suitable CCC. In this case, only one 'reference' ULCA is to be externally calibrated with a CCC while the other units may be calibrated against the reference with very low uncertainty. Furthermore, the setup allows performing ULCA stability investigations with very high accuracy [18].

## VII. Summary and Conclusions

The setup using an ULCA calibrator as current source, shown in Fig. 1, was verified to be suitable for the calibration of electrometers with sub-ppm accuracy, and for studying the time dependence of relevant DUT parameters. The ULCA calibrator's fast settling and its low $1/f$-corner enable investigating electrometers with respect to the time dependence of their burden voltage and settling behavior between $\approx 1$ s and $\approx 1000$ s. This excels the possibilities of the "voltage-ramp-to-capacitor" method [19], which to date still is used for electrometer calibrations [20–23]. Accuracy limits of the setup for calibrations in the current range between 1 fA and 1 µA were discussed.

As the comparison of the measurement results of the four commercial electrometers investigated shows, DUT4 (ULCA-1) shows unparalleled performance with respect to gain factor stability, settling and burden voltage performance, and temperature coefficient.

Also, a further advanced setup for the calibration of transimpedance amplifiers was presented. It uses the same DVM for measuring the output voltages of the ULCA calibrator and of the electrometer to be calibrated (given that its transimpedance is of the level of the ULCA calibrator) and, thus, eliminates DVM gain and linearity uncertainties. This setup enables calibrations of ULCA instruments without use of a cryogenic current comparator, but at similar accuracy level and in reasonably short time.

Altogether, the results verify the ULCA as an excellent instrument for small-current calibrations, and show that corresponding methods set new benchmarks on the accuracy limits for electrometer calibrations in the current range between 1 pA and 1 µA: with the setup shown in Fig. 1, sub-part-per-million accuracy is possible for "ULCA-ULCA" calibrations, while calibrations using the advanced setup shown in Fig. 8 can even be performed at accuracy better than 0.1 µA/A. Corresponding impact on related calibration and measurement capabilities can be expected from a comparison with PTB's present CMCs [16]. Regarding future comparisons in small-current metrology, the ULCA is predestined for being used as robust traveling standard [3, 18].

## Acknowledgment

The authors thank F. J. Ahlers and B. Schumacher (both with PTB) for valuable discussions.